\documentclass[fleqn,usenatbib]{mnras}

\usepackage[T1]{fontenc}
\usepackage{comment}
\DeclareRobustCommand{\VAN}[3]{#2}
\let\VANthebibliography\thebibliography
\def\thebibliography{\DeclareRobustCommand{\VAN}[3]{##3}\VANthebibliography}

\usepackage{graphicx}	
\usepackage{amsmath}	
\usepackage{caption}
\usepackage{subcaption}
\usepackage{booktabs}
\usepackage{makecell}
\usepackage{multirow}
\usepackage{hyperref}
\usepackage{amssymb}	
\usepackage{blindtext}
\usepackage{txfonts}

\title[Broadband study of the Blazar Ton\,599 ]{Broadband spectral and temporal study of Ton\,599 during the brightest January 2023 flare}

\author[Aaqib Manzoor et al.]{
Aaqib Manzoor$^{1}$\thanks{E-mail: aqibmanzoor1111@gmail.com}, 
Zahir Shah$^{2}$\thanks{E-mail: shahzahir4@gmail.com}, Sunder Sahayanathan$^{3,4}$,  
 Naseer Iqbal$^{1}$ \& Athar A. Dar $^{1,2}$  \\
$^{1}$Department of Physics, University of Kashmir, Srinagar 190006, India.\\
$^{2}$Department of Physics, Central University of Kashmir, Ganderbal 191201, India.\\
$^{3}$Astrophysical Sciences Division, Bhabha Atomic Research Center, Mumbai 400085, India.\\
$^{4}$Homi Bhabha National Institute, Mumbai 400094, India.
}

\date{Accepted XXX. Received YYY; in original form ZZZ}

\pubyear{2015}

\begin{document}
\label{firstpage}
\pagerange{\pageref{firstpage}--\pageref{lastpage}}
\maketitle

\begin{abstract}
In this work, we provide a detailed analysis of the broadband temporal and spectral properties of the blazar Ton\,599 by using the observations from \emph{Fermi}-LAT and  \emph{Swift}-XRT/UVOT telescopes, during its brightest $\gamma$-ray flaring. The one-day bin $\gamma$-ray light curve exhibits multiple substructures with asymmetric and symmetric profiles. Notably,  the $\gamma$-ray light curve shows a maximum flux of $\rm 3.63 \times 10^{-6}\, ph \,cm^{-2}\,s^{-1}$ on MJD\,59954.50, which is the highest flux ever observed from this source. The correlation between the $\gamma$-ray flux and $\gamma$-ray spectral indices suggests a moderate harder when the brighter trend. Taking $\gamma$-ray light curve as the reference, a strong correlation is observed with X-ray, optical, and UV energies. Additionally,  the $\gamma$-rays and optical/UV emissions exhibit higher variability compared to X-rays. To understand the parameter variation during the active state of the source, we conducted a statistical broadband spectral modelling of the source in 10 flux intervals of equal duration. A one-zone leptonic model involving synchrotron, synchrotron-self-Compton, and external-Compton processes successfully reproduces the broadband SED in each of these flux intervals. 
We observed that the flux variation during the active state is mainly associated with the variation in the magnetic field and the particle spectral indices.  

\end{abstract}




\section{Introduction}
Blazars are radio-loud Active Galactic Nuclei (AGN) with a powerful relativistic jet inclined close to the observer's line of sight. These are powered by a supermassive black hole (SMBH) located at their centre \citep{1995PASP..107..803U}.  The small angle of inclination of the relativistic jet induces  Doppler boosting in blazar emission, leading to distinctive observational features in all the energy bands \citep{2013ApJ...768...54B, 1997ARA&A..35..445U}.
The temporal variability of blazars extends from years \citep{2013MNRAS.436.1530R} to minutes \citep{2007ApJ...664L..71A,2007ApJ...669..862A}. Blazars are categorized into two main classes: Flat Spectrum Radio Quasars (FSRQs) and BL Lacertae (BL\,Lacs) objects, 
FSRQs are characterised by prominent emission line features, whereas BL\,Lacs typically display either weak or no emission line features \citep{1995PASP..107..803U}.

Blazars display a distinctive broadband spectral energy distribution (SED) marked by two prominent broad peaks. The first peak is typically observed in the Optical/UV/X-ray energy spectrum, while the subsequent peak is observed in the $\gamma$-ray energy range \citep{1998MNRAS.299..433F,2016ApJS..224...26M}. The lower energy peak results from the synchrotron cooling of electrons within the blazar jet, while investigations into the precise mechanisms governing the high energy emission are still ongoing. The high energy peak in the SED is generally associated with the inverse-Compton (IC) scattering \citep{2010ApJ...716...30A}. 
 The seed photons for IC scattering can be the synchrotron photons from the jet, and IC scattering of synchrotron photons is described as synchrotron-self-Compton process \cite[SSC:][]{1974ApJ...188..353J,1985A&A...146..204G,2002A&A...384...56C}). They can also originate from outside the jet, and IC scattering of these photons is described as an external-Compton process \cite[EC:][] {1992A&A...256L..27D,1994ApJ...421..153S,2000ApJ...545..107B,Shah_2017}).
The outside photon sources primarily include the broad line region \cite[BLR:][] {1996MNRAS.280...67G}, dusty torus \citep{2000ApJ...545..107B,2008MNRAS.387.1669G} and accretion disk \citep{1993ApJ...416..458D,1997A&A...324..395B}. In some cases, the high energy component of blazar SED is explained with a hadronic model through proton synchrotron process and/or by pion decay processes \citep{1992A&A...253L..21M,1993A&A...269...67M,2001APh....15..121M,2003APh....18..593M}. Generally, the preference for the leptonic model over the hadronic model typically stems from the considerable jet power demanded by the latter.
\citep{2013ApJ...768...54B,2016ApJ...825L..11P}. The peak synchrotron frequency ($\nu_{syn}^p$) for blazars in the $\nu-\nu f_{\nu}$ plot is an important observational feature that distinguishes FSRQs from BL\,Lacs \citep{1998MNRAS.299..433F}. For FSRQs $\nu_{syn}^p$ ranges between $10^{12.5}$ and $10^{14.5}$ Hz, while in BL\,Lacs its value is between $10^{13}$ and $10^{17}$ Hz   \citep{2010ApJ...716...30A}.
 BL\,Lacs are categorised into three groups depending on the position of $\nu_{\text{syn}}^p$ within the SED: low-energy peaked BL\,Lacs (LBL) if $\nu_{\text{syn}}^p$ is below $10^{14}$ Hz, intermediate-energy peaked BL\,Lacs (IBL) if $\nu_{\text{syn}}^p$ falls between $10^{14}$ Hz and $10^{15}$ Hz, and high-energy peaked BL\,Lacs (HBL) if $\nu_{\text{syn}}^p$ exceeds $10^{15}$ Hz \citep{2010ApJ...716...30A}.

Ton\,599, identified as a bright  FSRQ, is situated at a redshift $z \sim 0.725$ \citep{2010MNRAS.405.2302H}. Positioned at equatorial coordinates R.A = 179.883 and Dec. = 29.2455 degrees, it was first detected by the Energetic Gamma Ray Experiment Telescope (EGRET) in $\gamma$-ray energy \citep{1995ApJS..101..259T}. Following the launch of the \emph{Fermi}-LAT satellite, Ton\,599 was detected within the first three months \citep{2010Natur.463..919A}. Remarkably, it stands out as one of the few FSRQs detected in TeV energies \citep{2017ATel11075....1M,2017ATel10817....1M}. This source has been also observed in other energy bands over the past three decades, with a primary focus on investigating the flux variability. The SSC model was employed in the early 1980s to explain the radio and $\gamma$-ray emission from the source \citep{1983ApJ...274..101G}. The study carried out by  \citet{1993MNRAS.261..464M} on Ton\,599 suggests that the relativistic jet which initially is inclined at a very small angle with respect to the observer undergoes a small bending at later stages.  \citet{2014MNRAS.445.1636R} carried out a correlation analysis between the radio and $\gamma$-ray energy bands with the aim of  constraining the size of the $\gamma$-ray emitting region.
Several studies, such as  \citet{2008MNRAS.385..283C, Ghisellini_2014,2017ApJ...851...33P}  have employed a one-zone model to perform broadband SED modelling of the source. The primary objective of these studies was to explore the power of jets and other parameters of blazars in different flux intervals. 
In November 2017, the Ton\,599 underwent a prolonged flare for the first time across the electromagnetic spectrum. During this flare, the $\gamma$-ray flux  reached the maximum of $ \rm 1.26\times 10^{-6} ph \,cm^{-2} s^{-1}$ \citep{Prince_2019}. A detailed multi-frequency variability study of this flare was carried out by \citet{Prince_2019}, while the same flare was modelled using the leptonic model by \citet{2020MNRAS.492...72P}.
 They constrained the origin of $\gamma$-rays to be outside the BLR region and estimated the size of the emission region as  $\sim 1.03 \times 10^{16}$ cm. Recently, the Ton\,599 displayed its brightest ever $\gamma$-ray flare in January 2023. This flare was simultaneously observed in optical/UV and X-ray energy bands by \emph{Swift}-UVOT and \emph{Swift}-XRT, thereby allowing us to carry out a detailed multi-wavelength temporal and spectral study of this flaring period.

 In this work, we carry out a detailed multi-wavelength temporal and spectral study of Ton\,599 for the January 2023 flare.  This paper consists of 5 Sections. In section~\ref{analysis}, we give details of the data analysis procedure and observations used in this work. In section~\ref{sec3}, we present the temporal properties of the source. In section~\ref{spectral}, we provide the spectral properties of the source. Finally, the summary and discussion of results are given in section~\ref{diss}.

\section{Observations and Data analysis}
\label{analysis}
To investigate the temporal and spectral characteristics of the January 2023 flare in Ton\,599, we utilized data from \emph{Swift}-Ultraviolet/Optical Telescope \citep[UVOT;][]{2005SSRv..120...95R},  X-ray telescope \citep[XRT;][]{2005SSRv..120..165B} and \emph{Fermi}-LAT \citep{2009ApJ...697.1071A} telescope. 
The following section provides details of the data analysis techniques and instruments utilized in this study.
 \subsection{\emph{Fermi}-LAT Analysis}
The \emph{Fermi}-LAT  is an integral component of the \emph{Fermi} $\gamma$-ray space telescope, a satellite launched by NASA in 2008. Operating within the energy range of 20 $\rm MeV$ to 300 $\rm GeV$, it is designed to detect and study $\gamma$-rays. It converts the $\gamma$-ray photons into electron-positron pairs. It possesses a broad field of view of around $\sim 2.3$ Sr.
 To study the January 2023 flare, we conducted an analysis using the \emph{Fermi}-LAT PASS 8 data obtained during the period MJD 59884--59992. This data is publicly available at HEASARC \footnote{https://fermi.gsfc.nasa.gov/cgi-bin/ssc/LAT/LATDataQuery.cgi}. We selected a circular area with a radius of $15\deg$ around the location of Ton\,599 as the region of interest (ROI). Photons gathered within the energy span of 0.1 -- 300 $\rm GeV$ are taken into consideration.
 The \emph{fermipy}-v1.0.1 \citep{2017ICRC...35..824W} and the instrument function (IRF) \emph{$P8R3_-SOURCE_-V3$} are used for analyzing the data. For analysis, we adhered to the standard procedure outlined in \emph{Fermi}-LAT documentation 
 \footnote{http://fermi.gsfc.nasa.gov/ssc/data/analysis/}. 
To take care of the earth limb contamination, the zenith angle was chosen to have a maximum value of 90 degrees. We acquired a model file that contains all the sources given in the \emph{Fermi}-LAT 4FGL catalogue. 
During analysis, the model parameters of all
sources which fall within ROI were kept free, whereas, for sources lying beyond ROI, the model parameters were frozen to their 4FGL catalogue values.
To account for galactic and isotropic emission contributions, \emph{$gll_-iem_-v07.fits$} and \emph{$iso_-P8R3_-SOURCE_-V3_-v1.txt$}  were added to the model file during fitting. 
The good time intervals (GTI) were selected using a recommended criteria of  ``($DATA_-QUAL>0)\&\&(LAT_-CONFIG==1)$''. For $\gamma$-ray spectral and light curve generation, source detection is considered only when test statistics (TS) is > 9  \citep[$\sim 3\sigma$ detection;][]{1996ApJ...461..396M}. In this work, we have obtained a one-day  binned $
\gamma$-ray light curve of Ton\,599 during the period MJD 59884–59992.  Additionally, we acquire the $\gamma$-ray spectra of the source during the simultaneous X-ray and optical/UV observations for the broadband spectral analysis.

\subsection{\emph{Swift}-XRT/UVOT}
\label{xrt,uvot}
The \emph{Swift} is a multi-wavelength satellite launched by NASA on the 20\emph{th} of, November 2004 to monitor transient events occurring in both the galactic and extra-galactic skies. It possesses a viewing angle of 23.6 $\times$ 23.6 arcmins. Featuring an effective area of 110 cm² and a resolution of 18 arcseconds, the satellite consist of three onboard telescopes \citep{2005SSRv..120...95R, 2005SSRv..120..165B}. 
The \emph{Swift} observed the January 2023 flare of Ton\,599 in Ultraviolet/Optical and X-ray energy bands. The data is publicly available at HEASARC \footnote{\url{https://heasarc.gsfc.nasa.gov/cgi-bin/W3Browse/w3browse.pl}}. The details of the observations carried out during the January 2023 flaring period of Ton\,599 are given in Table~\ref{tab:xrt_obs}.\\
In the case of \emph{Swift}-XRT,  the clean event files were generated by running \emph{xrtpipeline}, utilizing the calibration file (CALDB, version: 20190910). The \emph{xselect} package is employed for selecting the source and background regions, as well as for the subsequent generation of the spectra of these regions. The source region is defined as a circular area with a radius of 25 pixels, centred at the source position. Similarly, the background region is also circular, with a radius of 50 pixels, positioned in a region devoid of the source.
An auxiliary response file (ARF) is generated using \emph{xrtmkarf}. 
The spectra have been grouped using the \emph{grppha} tool, ensuring that each bin contains a minimum of 20 counts. Subsequently, the grouped spectra within the energy range of 0.3-10 keV were fitted using \emph{xspec} \citep{1996ASPC..101...17A}, a component of the HEASoft package (version 6.30.1). The spectra were fitted by employing an absorbed power-law model. During the spectral fitting, the neutral hydrogen column density $n_H$ is frozen at $\rm 1.63 \times 10^{20} \, cm^{-2}$ \citep{2005A&A...440..775K}, while other parameters are set free.\\
The \emph{Swift}-UVOT has also observed the January 2023 flare of Ton\,599 utilizing the available six filters.
The  \emph{uvotsource} package incorporated within HEAsoft (version 6.30.1) was used to analyze the UVOT data.
Multiple images within a filter were combined using the \emph{uvotsum} tool. Photons originating from the source were collected from a circular region of 5 arcsec radius, centred at the source position. In contrast, background photons were obtained from a circular region with a 10 arcsec radius, located in a source-devoid region.
 The multiple images in a filter are added with \emph{uvotsum} tool.  
The observed flux points are de-reddened for galactic extinction, by the procedure outlined by \citet{2011ApJ...737..103S}.

\begin{table*}
\centering
\caption{Details of \emph{Swift}-XRT/UVOT observations.}
\begin{tabular}{l c c c  c}
\hline 
S. No. & Observation ID & Time (MJD) & XRT exposure (ks) & UVOT exposure (ks)\\
\hline
\hline
1.  & 00036381097 & 59943.5   & 1.41 & 1.38\\
2.  & 00036381098 & 59946.8   & 1.49 & 1.47\\
3.  & 00036381099 & 59949.4   & 1.01 & 0.98\\
4.  & 00036381100 & 59952.7   & 1.58 & 1.55\\
5.  & 00036381101 & 59955.1   & 1.59 & 1.52\\
6.  & 00036381102 & 59959.3   & 1.78 & 1.70\\
7.  & 00036381103 & 59960.4   & 3.82 & 3.75\\
8.  & 00036381104 & 59964.7   & 1.53 & 1.50\\
9.  & 00036381105 & 59967.4   & 1.62 & 1.60\\
10. & 00036381106 & 59970.4   & 1.67 & 1.65\\
\hline
\hline
\end{tabular}
\vspace{0.5cm}
\label{tab:xrt_obs}
\end{table*}

\section{Results}
\label{sec3}

\subsection{Temporal study}
\label{temporal}
\begin{figure*}

    \includegraphics[width=0.8\linewidth,height=.4\textheight]
    {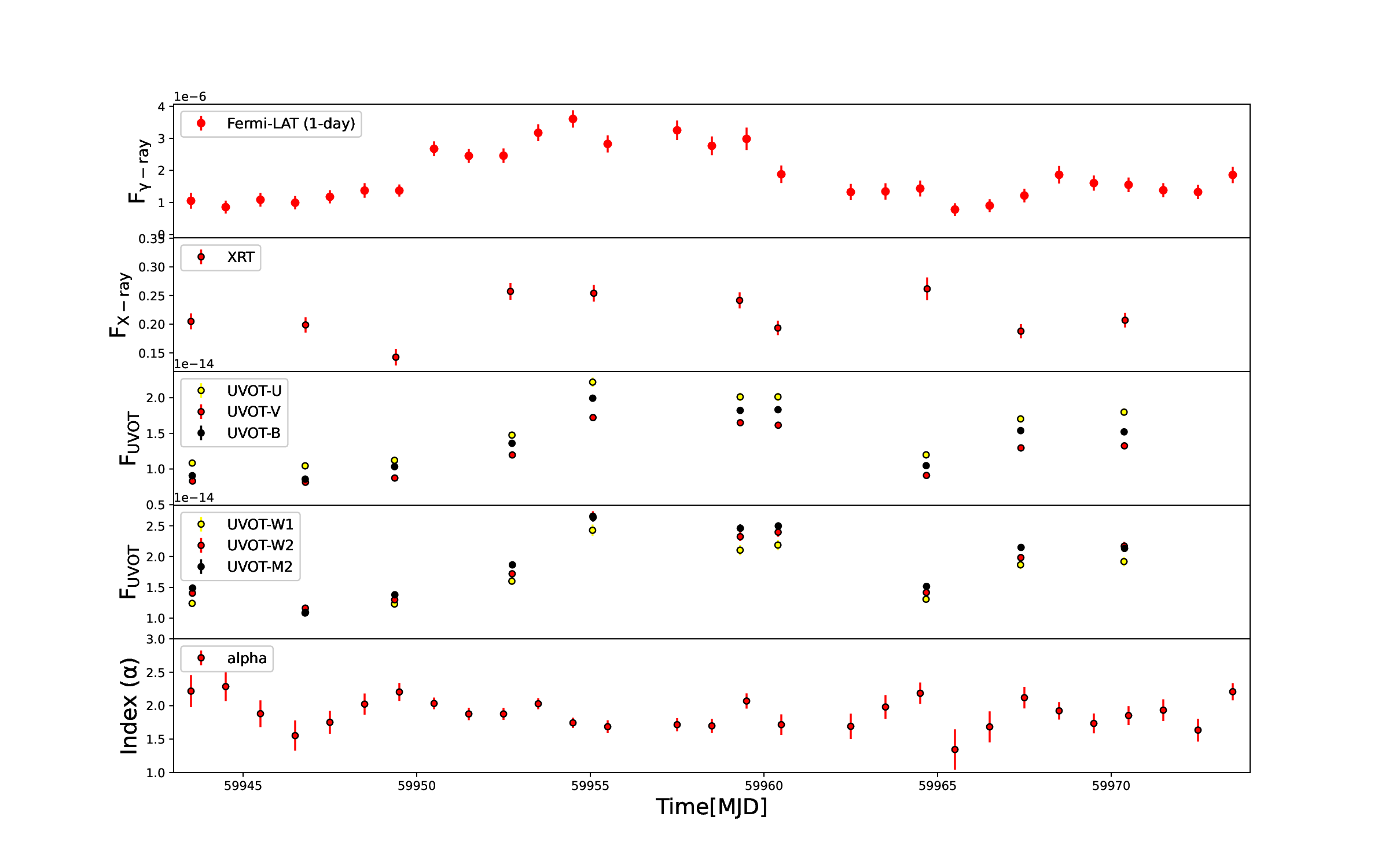}
    \caption{Multi-wavelength light curves for Ton\,599 during MJD 59943-59974, featuring (1) top panel:daily-binned $\gamma$-ray light curve (0.1--300 $\rm GeV$) from Fermi-LAT (with unit $ \rm ph \,cm^{-2}\,s^{-1}$) , (2) second panel: X-ray light curve (0.3--10.0 keV) from \emph{Swift}-XRT  (counts/sec), (3) third and fourth panel: Optical/Ultraviolet light curves from \emph{Swift}-UVOT in $\rm erg/cm^{2}/s$/Å.  The final panel presents the temporal evolution of spectral slope values.}

    \label{1_d}
\end{figure*}

\begin{figure*}

    \includegraphics[width=1.0\linewidth,height=.44\textheight]
    {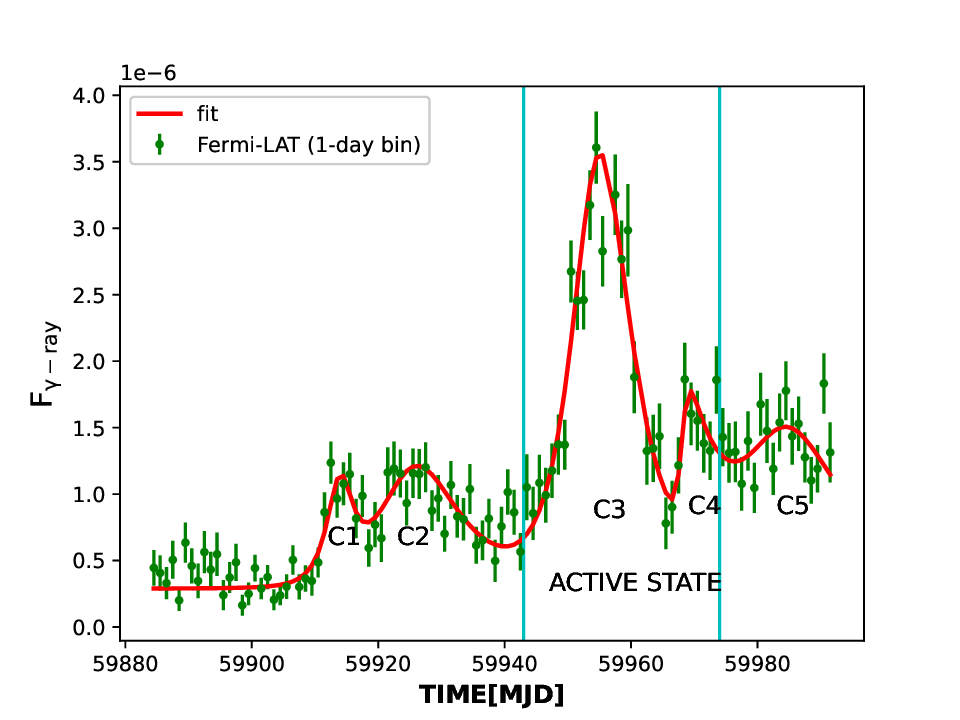}
    \caption{ The daily-binned $\gamma$-ray light curve of Ton\,599 obtained in the energy range of 0.1 -- 300 $\rm GeV$ during the period MJD 59884 -- 59992. The $\gamma$-ray flux is in units of $\rm ph cm^{-2}s^{-1}$. The solid red curve shows the best fit, which is the sum of exponential functions. The period delineated by two vertical cyan-coloured lines, corresponding to MJD 59943 -- 59974, represents the active state of the source.
}
        \label{new_component}

\end{figure*}

To investigate the temporal features of Ton\,599, we generated its multi-wavelength light curves (MWL LC) in UV/optical, X-ray, and $\gamma$-ray energy bands during the period MJD 59884 -- 59992. The daily binned $\gamma$-ray light curve is obtained by integrating photons within the 0.1 -- 300 $\rm GeV$. The Ton\,599 has been observed across multiple wavelengths during the period MJD 59943 -- 59974. Figure~\ref{1_d} displays the MWL LC of the Ton\,599 by using the \emph{Fermi} and \emph{Swift} observations. 
The base-flux in $\gamma$-ray light curve is obtained as $\rm 2.9 \times 10^{-7} \, ph \,cm^{-2} s^{-1}$, which is calculated by taking the average of fluxes in the quiescent state during the time interval MJD 59885 -- 59904. We define the active state of the source as the period where the $\gamma$-ray flux has enhanced significantly above the base flux and multi-wavelength observations are available. As shown in  Figure~\ref{new_component}, $\gamma$-ray flux during the time interval MJD 59943-59974 has increased significantly above the base-flux. We call this time interval an active state of the source. In the daily binned $\gamma$-ray light curve, the source has shown a peak flux of $\rm 3.63 \times 10^{-6}\,ph \,cm^{-2} s^{-1}$ on MJD 59954.50. This is the highest-ever flux reported from Ton\,599 till now. It is about  $\sim$ 13 times larger than the base flux of the source. A closer look at a one-day binned $\gamma$-ray light curve (see Figure~\ref{new_component})  shows that a light curve consists of five dominant component shape, we name them C1, C2, C3, C4, and C5. We fit these components by an exponential function of the form \citep{2010ApJ...722..520A}

\begin{equation}
    F(t) = F_b + \frac{2F_0}{\left[exp \left(\frac{t_0 - t}{T_r}\right) + exp \left(\frac{t - t_0}{T_d}\right)\right]}~~~,
    \label{func}
\end{equation}
to calculate the rise and decay time of these components. Here $F_b$ represents base flux,  $t_0$ is the peak time, $F_0$ is peak flux,  $T_r$ and $T_d$ represent the rise and decay times of the individual components, respectively. The fitted profile of the light curve is shown in Figure~ \ref{new_component}, and the best-fit parameters are shown in Table~\ref{rise_decay}. Additionally, we have calculated the asymmetric parameter ($\zeta$) for each component. The asymmetric parameter measures the strength of symmetry in each component. It varies between 0 -- 1 and is given by  $\zeta = \frac{T_d - T_r}{T_d + T_r}$. A component is said to be symmetric if $\zeta$ < 0.3, moderately symmetric if 0.3 < $\zeta$ < 0.7 and asymmetric if $\zeta$ > 0.7. The calculated values of $\zeta$ parameter are given in Table~\ref{rise_decay},  the results suggest that the components C1, C2, and C3 are symmetric, while the component C4 is asymmetric. However, due to substantial error in the asymmetric parameter value of component C5, its profile is categorized as both symmetric and moderately asymmetric.\\

\begin{table*}
\centering
\caption{Rise and decay times of different components of $\gamma$-ray light curve for Ton\,599 during  the time  MJD 59884--59992. Column: 1. Component, 2. Peak time (MJD), 3. Peak flux, 4.  Rise time  5. Decay time and 6. Asymmetry parameter.}
\begin{tabular}{c c c c c c} 
\hline 
Components & $\rm t_0$ & $\rm F_0$ & $\rm T_r$ &  $\rm T_d$ & $\rm |\zeta|$ \\
     &  (MJD) & ($\rm 10^{-6}\ ph\ cm^{-2}\ s^{-1}$) & (days) & (days) & \\
\hline
\hline
C1   &  59914 & 0.7 & 1.78 $\pm$ 0.43  & 1.69 $\pm$ 0.62 & 0.025$\pm$ 0.28\\

C2   &  59925 & 0.9 & 4.61$\pm$ 0.90  & 7.04 $\pm$ 0.62 & 0.20$\pm$ 0.09\\

C3   &  59954.44 & 3.2 & 3.49$\pm$ 0.17  & 4.86 $\pm$ 0.27 & 0.16$\pm$ 0.04\\

C4   &  59968.15 & 0.7 & 0.56$\pm$ 0.28  & 7.21$\pm$ 1.59 & 0.85$\pm$ 0.02\\

C5   &  59983 & 1.0 & 4.71$\pm$ 1.28  & 10.0$\pm$ 2..48 & 0.35$\pm$ 0.16\\
   
\hline
\hline
\end{tabular}
\vspace{0.5cm}
\label{rise_decay} 
\end{table*}

Following the detection of high $\gamma$-ray activity by \emph{Fermi}-LAT, \emph{Swift} conducted 10 observations of the Ton\,599 source during its active state (MJD 59943-59974), the description of these observations is given in Table~\ref{tab:xrt_obs}. We use these observations to investigate the behaviour of the source at optical, UV, and X-ray energy bands. The  X-ray, optical, and UV light curves are presented in the second, third, and fourth panels of the MWL LC respectively (Figure~\ref{1_d}).  
The X-ray data points are acquired within the energy range of 0.3 -- 10 keV, with each point in these light curves representing a specific observation ID. The visual inspection of MWL LC (see Figure~\ref{1_d}) suggests correlated flux variations within different energy bands. We look for the correlation between $\gamma$-ray flux and other energy bands by taking the $\gamma$-ray flux as a reference and utilizing the Spearman rank correlation method. The resulting $r_s$ and  $P_{rs}$  listed in Table~\ref{corr}  confirm the presence of correlated flux variations across distinct energy bands.\\

Blazars show variability in all the observed energy bands. These variations are prominent during flaring events. The variability amplitude depends upon source parameters \emph{viz} magnetic field, viewing angle, and particle density   \citep{2018A&A...617A..59K}. The availability of multi-wavelength observations in optical, ultraviolet, X-ray, and $\gamma$-rays for Ton\,599 facilitates a comparison of variability amplitudes across different energy bands. The fractional variability  was calculated by following \citep{2003MNRAS.345.1271V}

\begin{equation}
    F_{var}=\sqrt{\frac{S^2-\langle \sigma_\mathrm{err}^{2}\rangle}{\langle  F \rangle^2}} \hspace{0.3cm},
\end{equation}
where $S^2$, $\langle  F \rangle$, and $\langle \sigma_\mathrm{err}^{2}\rangle$ are flux variance, mean flux and mean square error in the measurement of flux respectively. The uncertainty in the $F_\mathrm{var}$ is estimated using the formalism described in \citet{2008MNRAS.389.1427P}
\begin{equation}
\Delta F_\mathrm{var} = \sqrt[]{F_\mathrm{var}^2 + err(\sigma_\mathrm{NXS}^2)}-F_\mathrm{var}\hspace{0.3cm},
\end{equation}
where
\begin{equation}
err\left(\sigma_\mathrm{NXS}^2\right)=\sqrt{ \left( \sqrt{\frac{2}{N}} \cdot \frac{\langle \sigma_\mathrm{err}^{2}\rangle}{\langle F \rangle^{2}} \right)^{2} + \left(\sqrt{\frac{\langle\sigma_\mathrm{err}^{2}\rangle}{N}} \cdot \frac{2 F_\mathrm{var}}{\langle F\rangle} \right)^{2}} \hspace{0.3cm},
\end{equation}
where $N$ is the total number of flux points in the light curve. The calculated values of $F_{var}$ and their corresponding uncertainties $\Delta F_\mathrm{var}$ for optical, ultraviolet, X-ray, and $\gamma$-ray bands are given in Table~\ref{var}. As shown in the plot between $F_{var}$ and energy (see Figure~\ref{fig:frac_var}), the $F_{var}$ shows a dip at X-rays and then increases again towards the GeV band. This dip at X-rays has been also reported by many authors in other blazar sources  \citep{2016A&A...590A..61C,2016ApJ...819..156B,2021MNRAS.504..416S,2021MNRAS.504..416S,Malik_2022}. {The X-ray dip is linked to FSRQs' SED shape: high-energy electrons dominate optical/UV and $\rm \gamma-ray$ emissions via synchrotron and IC losses, respectively,  while X-rays result from IC losses due to the low-energy electrons. Since the highly energetic electrons cool faster, this results in a larger $F_{var}$ at optical/UV and GeV bands than at the X-ray band.

\begin{table}
\caption{The Spearman rank correlation results obtained by correlating the $\gamma$-ray flux (upper panel)  and the spectral index ($\alpha$) (lower panel) with different energy band flux. 
Col. 1: Correlating quantities, 2: Correlation coefficient value 3:  Null hypothesis probability value.}
\begin{tabular}{lcr}
\hline
Light curves & $r_s$  & $P_{rs}$ \\                 
\hline
$\gamma$-ray vs X-ray & 0.503 &  $0.138$         \\
$\gamma$-ray vs V     &  0.84 &  $0.002$         \\
$\gamma$-ray vs B     &  0.770 &   $0.009$        \\
$\gamma$-ray vs U    &  0.806 &   $0.004$          \\
$\gamma$-ray vs W1  &  0.770 &   $0.009$            \\
$\gamma$-ray vs M2  &  0.733  &   $0.015$            \\
$\gamma$-ray vs W2  & 0.770  &   $0.009$              \\
           
\hline
Index($\alpha$) &  $r_s$  & $P_{rs}$ \\
\hline
$\alpha$ vs $\gamma$-ray  & -0.278 &   0.008  \\
$\alpha$ vs X-ray  &    -0.115 &   0.751           \\
$\alpha$ vs V       &  -0.358  &  0.310 \\
$\alpha$ vs B      &   -0.358, &  0.310 \\
$\alpha$ vs U      &  -0.394  & 0.260 \\
$\alpha$ vs W1  &  -0.382 &   0.276  \\
$\alpha$ vs M2 &   -0.345 &   0.328  \\
$\alpha$ vs W2   &  -0.382 &   0.276  \\
\hline
\end{tabular}

\label{corr}
\end{table}

\begin{table}
\caption{ The calculated fractional variability amplitude ($F_{var}$) of \mbox{Ton\,599} in $\gamma$-ray, X-ray and optical/UV energy bands (upper panel) during the time period MJD 59884--59992. The variability of index ($\alpha$) is given in the lower panel.}
\begin{tabular}{lr}
\hline
Energy band  & $F_{var}$ \\
\hline
$\gamma$-ray (\mbox{0.1--300 $ \rm GeV$}) & 0.62$\pm$0.01  \\
X-ray (0.3--10 keV) & 0.17$\pm$0.02  \\
UVW2 & 0.32$\pm$0.01 \\
UVM2 & 0.32$\pm$0.01\\
UVW1 & 0.31$\pm$0.01\\
U & 0.32 $\pm$ 0.01\\
B & 0.34 $\pm$ 0.008\\
V & 0.33 $\pm$ 0.01\\
\hline
\hline
Index  & variability\\
Alpha($\alpha$) & 0.09 $\pm$ 0.02\\
\hline

\end{tabular}
\label{var}
\end{table}

\begin{figure}
		\centering
		\includegraphics[scale=0.5]{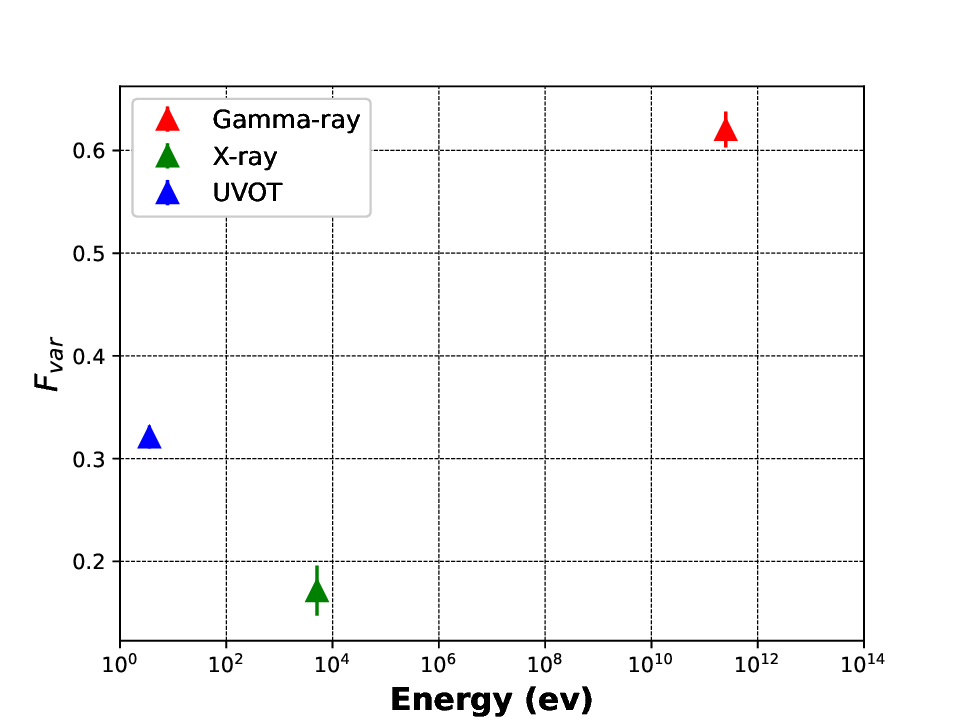}
		\vspace{0.5cm}
		\caption{The fractional variability amplitude ($\rm F_{var}$) of Ton\,599 computed across different energy bands and plotted against energy over the time  MJD 59884 -- 59992.}
		\label{fig:frac_var}
\end{figure}

\subsection{The gamma-ray spectral behaviour}
The one-day binned $\gamma$-ray light curve (see Figure~\ref{new_component}) has been obtained by fitting the spectrum (0.1 -- 300 $\rm GeV$) with a log-parabola model defined as
\begin{equation}
\frac{dN}{dE}=N_0\left(\frac{E}{E_p}\right)^{-\alpha-\beta\log\left(\frac{E}{E_p}\right)},
\label{LP}
\end{equation}
where $\alpha$ is the index at the pivot energy $E_p$, and $\beta$ is the curvature parameter. 
The index values obtained by fitting the daily averaged $\gamma$-ray spectrum, are plotted with time in the lower panel of  Figure~\ref{1_d}. The MWL LC suggests a correlated trend between the $\gamma$-ray flux and index values. Consequently, we carried out a correlation analysis between the $\gamma$-ray flux and the spectral index to explore the potential existence of the "harder when brighter" feature in Ton\,599.
 The  "harder when brighter" feature is a common feature observed in most of the blazars \citep{2016ApJ...830..162B, Shah_2021}.  The Spearman correlation coefficient and the corresponding null hypothesis probability values are obtained as $r_s \sim -0.28$ and $P_{rs} \sim 0.008$, respectively.
  A correlation plot between spectral indices and flux is shown in Figure~\ref{index}. The index versus flux points for various components are indicated by different colours. The colour scheme differentiates between the components with higher and lower flux. Notably, the components with brighter $\gamma$-ray flux exhibit a harder spectral index compared to those with lower flux. 
Overall, the results indicate a mild presence of the "harder when brighter" trend in Ton\,599. Additionally, we also calculated the correlation of $\gamma$-ray index with the X-ray and Optical/UV flux values, the correlation coefficient and null-hypothesis probability values are summarized in Table~\ref{corr}.  The obtained $P_{rs}$ values suggest no significant correlation between $\gamma$-ray index values and X-ray/Optical-UV flux values. However, it is essential to interpret these results cautiously, given the sparse nature of the data in X-ray and optical-UV bands.

\begin{figure}

		\begin{center}

	\includegraphics[angle=0,width=.45\textwidth]{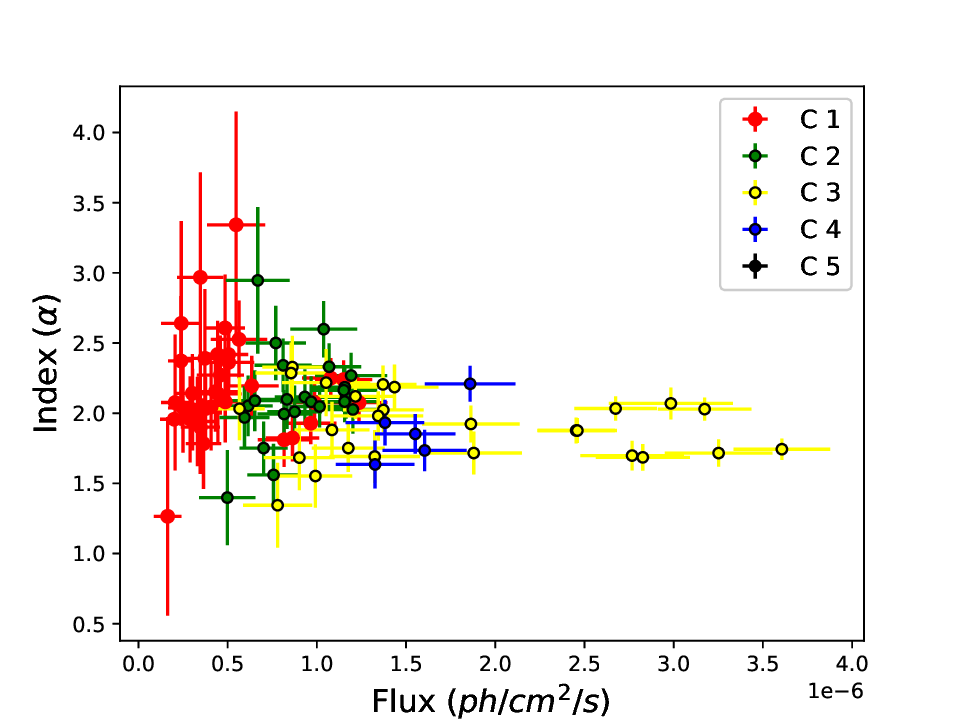}

        \caption{The Variation of spectral slope ($\alpha$) with $\gamma$-ray flux for Ton\,599 during the period MJD 59884 -- 59992. The C 1, C 2, C 3, C 4 and C 5 are the components of the $\gamma$-ray light curve (See Figure~\ref{new_component})}.
        \label{index}
		\end{center}        
\end{figure}

\section{Broadband Spectral Analysis}
\label{spectral}
In this section, we examine the fluctuations in the fundamental physical parameters responsible for the enhancement of flux during the active state (MJD 59943 -- 59974) by carrying out a detailed broadband spectral study of Ton\,599 using the $\gamma$-ray, X-ray, and optical/UV data. During the active state, the source has been observed 10 times by \emph{Swift} in X-ray and optical/UV energies. We divide the MWL LC (see Figure~\ref{1_d}) into ten flux intervals such that each state contains one \emph{Swift} observation. The duration of each flux state is chosen to be 24 hours with centre at \emph{Swift} observation. 
To obtain the $\gamma$-ray SED points for each state, we divided the total energy range 0.1 -- 300 $\rm GeV$ into eight equally spaced bins using a logarithmic scale. For the fitting of the $\gamma$-ray spectra, we employed the log parabola model. In our analysis, the parameters of Ton\,599 were allowed to vary freely, while the parameters of other sources within the Region of Interest (ROI) were kept frozen at their best-fit values obtained from fitting the $\gamma$-ray spectrum integrated over the energy range of 0.1 -- 300 $\rm GeV$.
 The X-ray spectra in different flux intervals are generated by using  \emph{xrtpipeline}, \emph{xselect} and \emph{ximage} tools (see Section~\ref{xrt,uvot} for details). We considered the  cpflux values in order  to account for the nH absorption in the X-ray spectra.
  In \emph{Swift}-UVOT, the images of each observation ID in different flux intervals are added together using \emph{uvotsum} tool and optical/UV spectral points are obtained from combined images. The obtained broadband spectral points for the selected intervals are plotted in Figure~\ref{fig:sed}. \\

The broadband SEDs for specific flux intervals are modelled using a one-zone leptonic model. \citep{2017MNRAS.470.3283S,Sahayanathan_2018}. 
We consider the emission to be originating from a spherical region of radius `R' within a relativistic jet characterized by a bulk Lorentz factor '$\Gamma$'.  The Jet is assumed to make a small angle `$\theta$' to the line of sight of an observer, resulting in the Doppler boosting of the observed flux. To incorporate the Doppler boosting effect on flux, we introduce the relativistic Doppler factor ($\delta$), defined as $\delta = [\Gamma (1 - \beta \cos \theta)]^{-1}$.
We assume that the spherical emission region is filled with electrons which are  distributed in energy as broken power-law (BPL), given by 
  \begin{align}
\label{eq:broken}
N(\gamma) d\gamma =\left\{
	\begin{array}{ll}
K \gamma^{-p}d\gamma,&\mbox {~$\gamma_{{\rm min}}<\gamma<\gamma_b$~} \\
K \gamma^{q-p}_b \gamma^{-q}d\gamma,&\mbox {~$\gamma_b<\gamma<\gamma_{{\rm max}}$~} 
\end{array} \quad ,
\right.
\end{align}
where $\gamma$ is the electron Lorentz factor, K is normalization, $\gamma_b$ is the spectral break energy, p and q are the particle spectral indices before and after the break. The observed flux is ascribed to synchrotron emission, a consequence of the presence of a magnetic field (B) and relativistic electrons, as well as IC emission, which includes SSC and EC processes.  Isotropic blackbody external photon fields for EC process with temperatures  T$\sim$1000 K and T$\sim$42000 K represent the IR torus and BLR photons, respectively. In FSRQs, high-energy emission involves contributions from both SSC and EC processes \citep{2017MNRAS.470.3283S}. The spectra are calculated from synchrotron, SSC and EC emissivity functions using a numerical code which is incorporated as a local model in  \emph{"xspec"}. The resultant local model is utilized to perform the broadband spectral fitting of the source. The main model parameters include \emph{viz}. $p$, $q$, $\gamma_b$, $B$, $R$, $U_e$, $\Theta$, $f$, $\Gamma$ and $T_*$, which are adjusted to achieve the best-fit SED. During the broadband SED fitting, we consider the equipartition condition between the particle energy density and magnetic energy density. 
Additionally,  a steady-state emission is assumed in the chosen flux intervals. 
To account for uncertainties associated with the model, we incorporated a 12\% systematic uncertainty into the data during the fitting of all the SEDs.
This inclusion of a 12\% systematic error enabled us to achieve a reduced $\chi^2$  value below 2 for each of the SEDs. Due to the limited data availability at $\gamma$-ray, X-ray, and optical/UV energies, we carry out broadband SED fitting by keeping $p$, $q$, $\Gamma$, and $B$ as free parameters, while the other parameters are fixed at typical values needed to get best-fit. The observed spectral points and the broadband SED fit along with the the contributions from synchrotron, SSC and EC components are shown in Figure~\ref{fig:sed}. The best-fit parameters are given in Table~\ref{parameters}.

\begin{table*}
	\centering
\caption{ Details of the best-fit parameters acquired in different flux intervals using one zone model. Free parameters:  Col. 1. Flux states; 2, low energy particle index 3. high energy particle index,  4. Bulk Lorentz factor, 5. Magnetic field (Gauss). Fixed parameters: Col. 6. Lorentz factor corresponding to break energy; 7. Minimum Lorentz factor of electrons; 8. Integrated $\gamma$-ray flux (in units $ 10^{-6} \,\rm ph \,cm^{-2}\,s^{-1}$), 9. $\chi^{2}$/degrees of freedom. Subscripts and superscripts on the values denote the lower and upper errors respectively, while "$--$" indicates that the error bound is not constrained. In all the SEDs,  $\gamma_{max} = 10^{6}$,  R = $10^{16}$ cm, T = 1000 K.} 
\vspace{0.5cm}
\begin{tabular}{c c c c c c c c c}
\hline
\hline

 State & p & q & $\Gamma_b$ & B & $\gamma_{b}$ & $\gamma_{min}$ & $F_{0.1-300 GeV}$  & $\chi^{2}$/dof  \\ 
 \hline
 \\
  1& $2.20^{+0.15}_{-0.20}$& $4.42^{+0.27}_{-0.20}$& $17.25^{+8.86}_{-4.97}$& $ 1.52^{--}_{-0.14}$ & 2700 & 150 & 1.05 &3.80/14  \\
  \\
   2& $2.34 ^{+0.11}_{-0.12}$& $4.92 ^{+1.09}_{-0.41}$& $38.94^{--}_{-21.29}$& $1.54^{--}_{-0.09}$ & 2400 & 140 & 0.99& 14.92/16  \\
\\
3& 2.01$ ^{+0.14}_{-0.21}$& $ 4.80^{+0.73}_{-0.48}$& $ 18.08^{+4.07}_{-2.91}$& $1.61^{--}_{-0.12}$& 2000 & 150  & 1.37&3.78/12\\

\\
4& $ 2.10^{+0.15}_{-0.72}$& $4.04^{+0.11}_{-0.11}$& $28.95^{+10.56}_{-8.20}$& $1.61^{+0.11}_{-0.09}$&1300 & 300& 2.46&10.63/14 \\
\\

5& $ 1.90^{+ 0.09}_{-0.09}$& $4.74^{+0.24}_{-0.20}$& $28.09 ^{+6.34}_{-4.67}$& $1.66^{+0.05}_{-0.03}$& 1950 & 160&2.82&21.85/15 \\
\\

6& $ 1.93^{+0.07}_{-0.09}$& $4.69^{+0.16}_{-0.20}$& $28.8^{+6.78}_{-5.49}$& $1.64 ^{+0.05}_{-0.02}$& 1900& 180& 2.98 &10.91/15 \\
\\
7& $ 1.80^{+ 0.07}_{-0.12}$& $ 4.61^{+0.17}_{-0.15}$& $29.02^{+12.40}_{-9.82}$& $1.61 ^{+0.10}_{-0.02}$& 1800 & 180 &1.87 &10.23/15 \\

\\
8& $ 2.03^{+0.15}_{-0.16}$& $4.41^{+0.47}_{-0.27}$& $46.76^{--}_{-11.34}$& $ 1.63^{+0.13}_{-0.10}$&2100 & 100 &1.43& 4.93/14 \\

\\
9& $ 1.90^{+0.13}_{-0.12}$& $ 4.78^{+0.49}_{-0.26}$& $45.76^{--}_{-10.49}$& $1.58^{+0.10}_{-0.11}$& 2700 & 90 & 1.21& 4.33/ 14\\

\\
10& $ 2.05^{+0.06}_{-0.11}$& $4.80^{+0.19}_{-0.22}$& $28.65^{+13.56}_{-9.18}$& $1.57^{+0.10}_{-0.02}$&2100 & 160 &1.55 & 6.39/16\\

\\
\hline
\hline
\end{tabular}
\label{parameters}
\end{table*}

\begin{figure*}
\vbox{
\hbox{
\includegraphics[scale=0.24, angle=270]{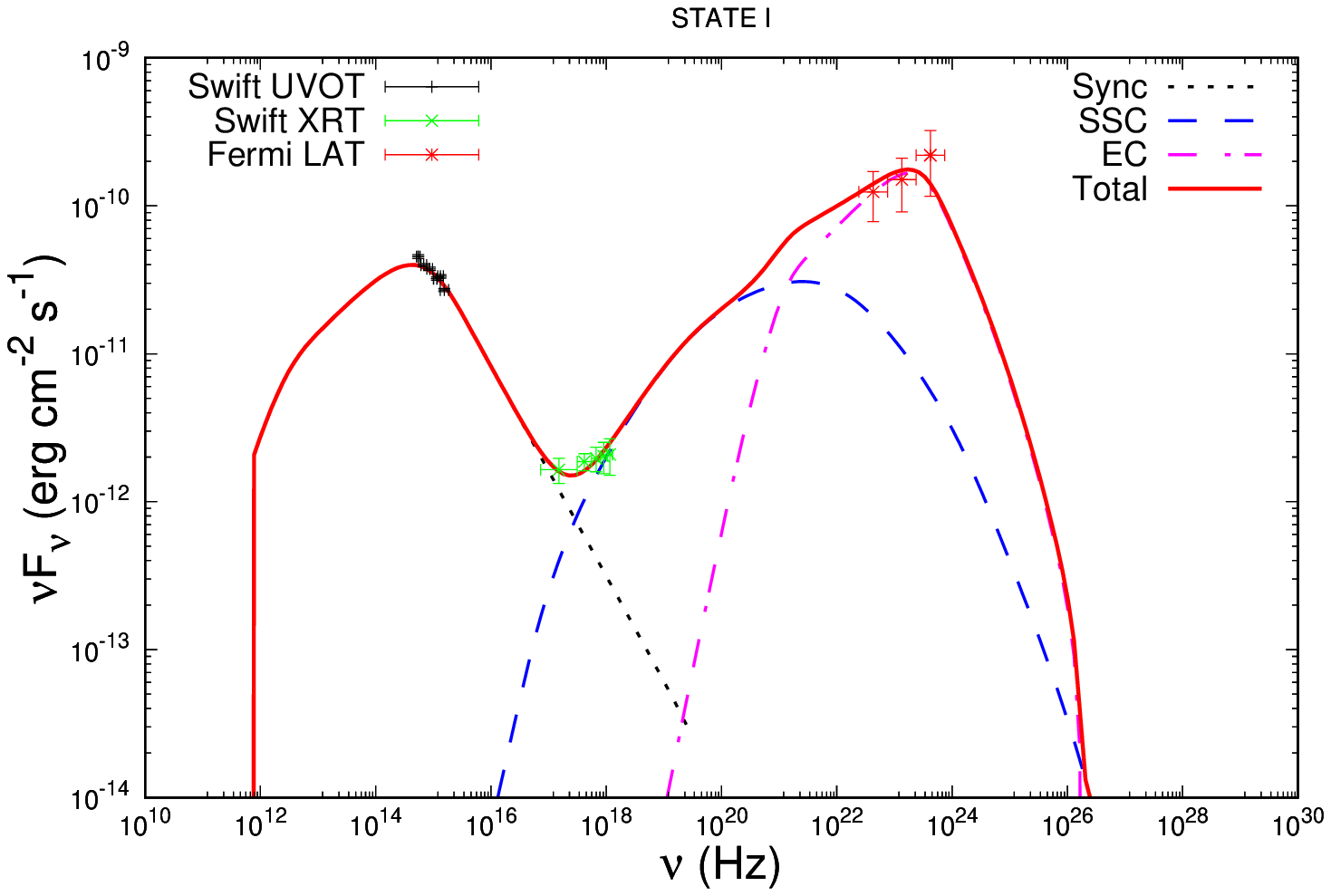}
\includegraphics[scale=0.24, angle=270]{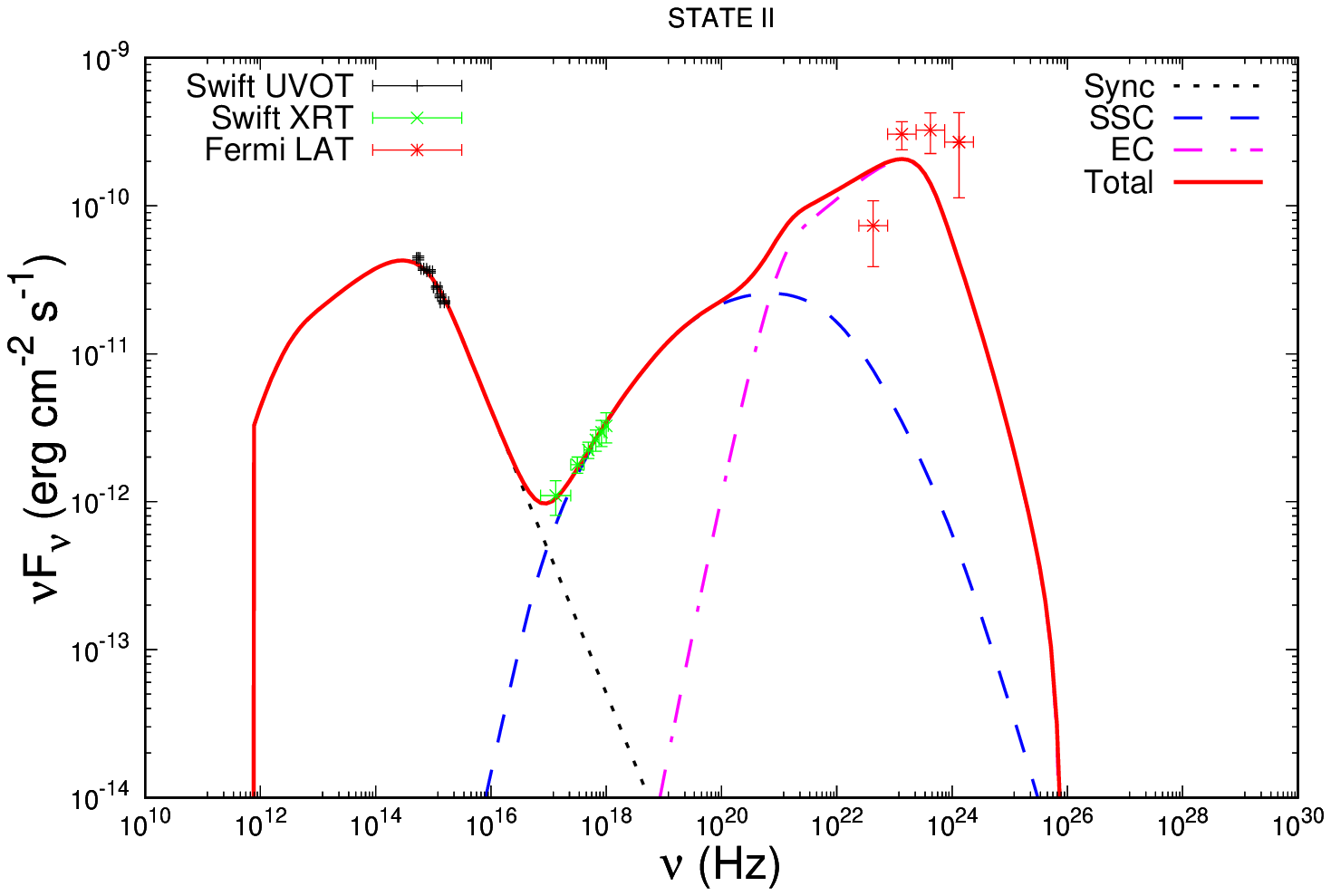}
\includegraphics[scale=0.24, angle=270]{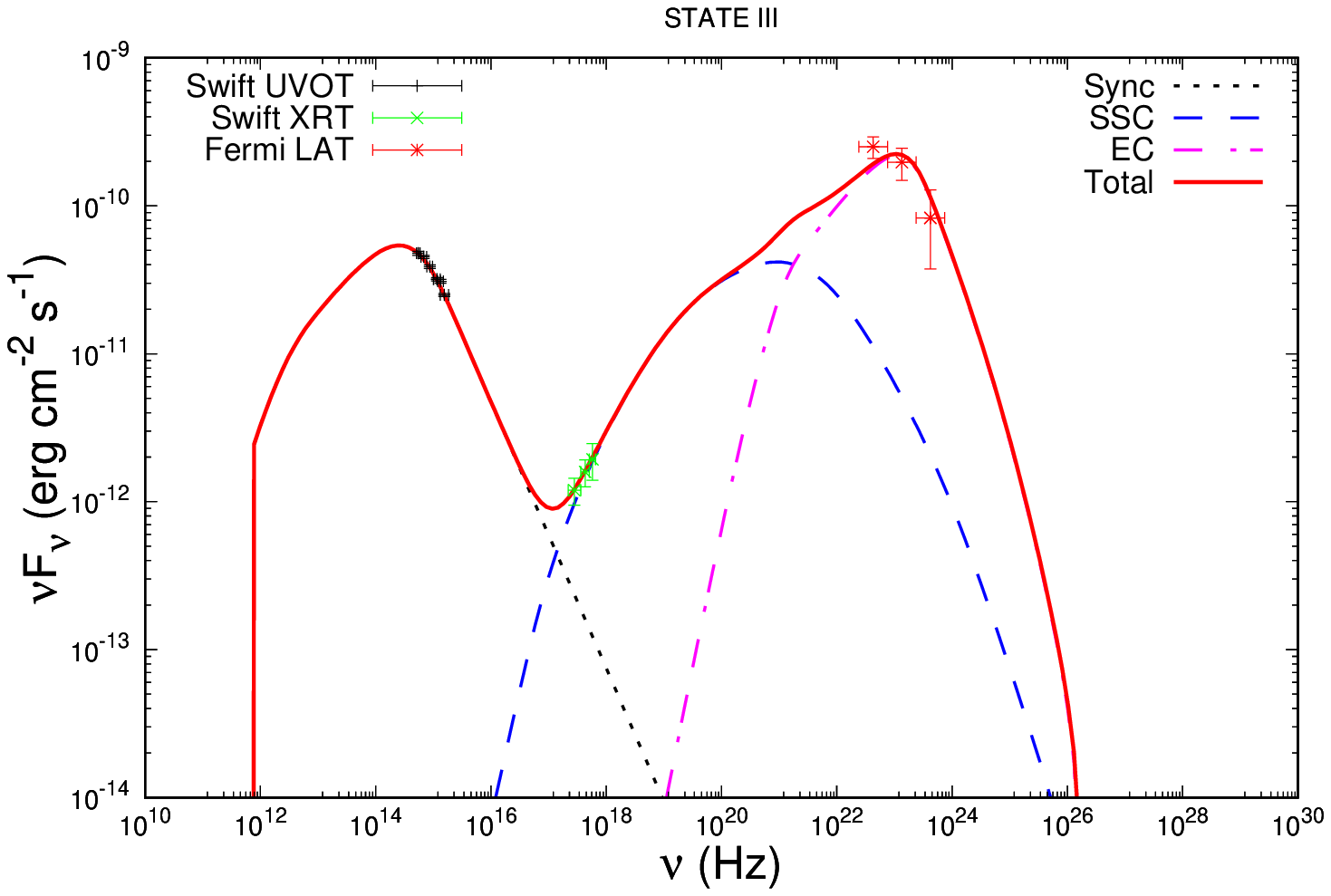}
    }
\hbox{
\includegraphics[scale=0.24, angle=270]{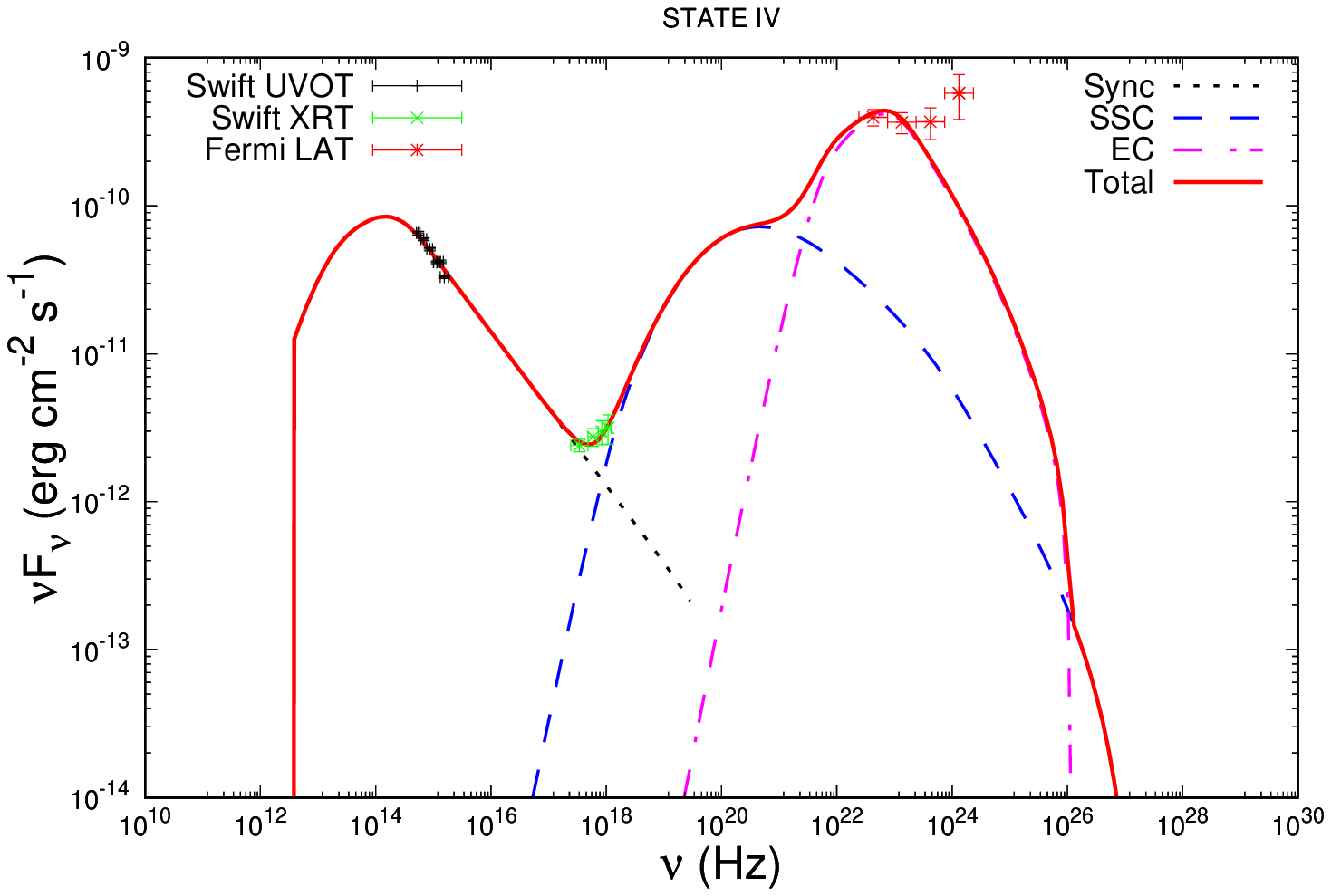}
\includegraphics[scale=0.24, angle=270]{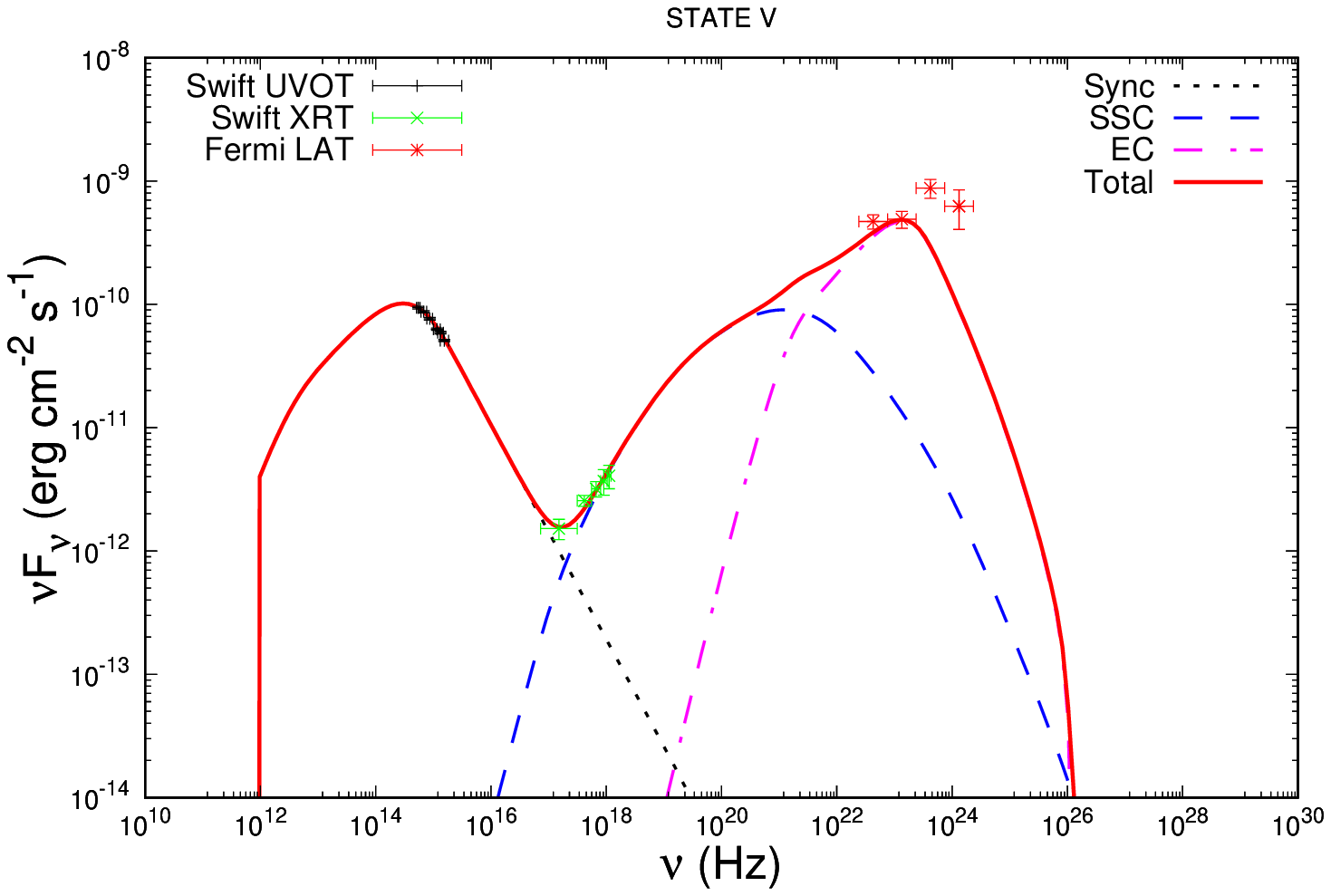}
\includegraphics[scale=0.24, angle=270]{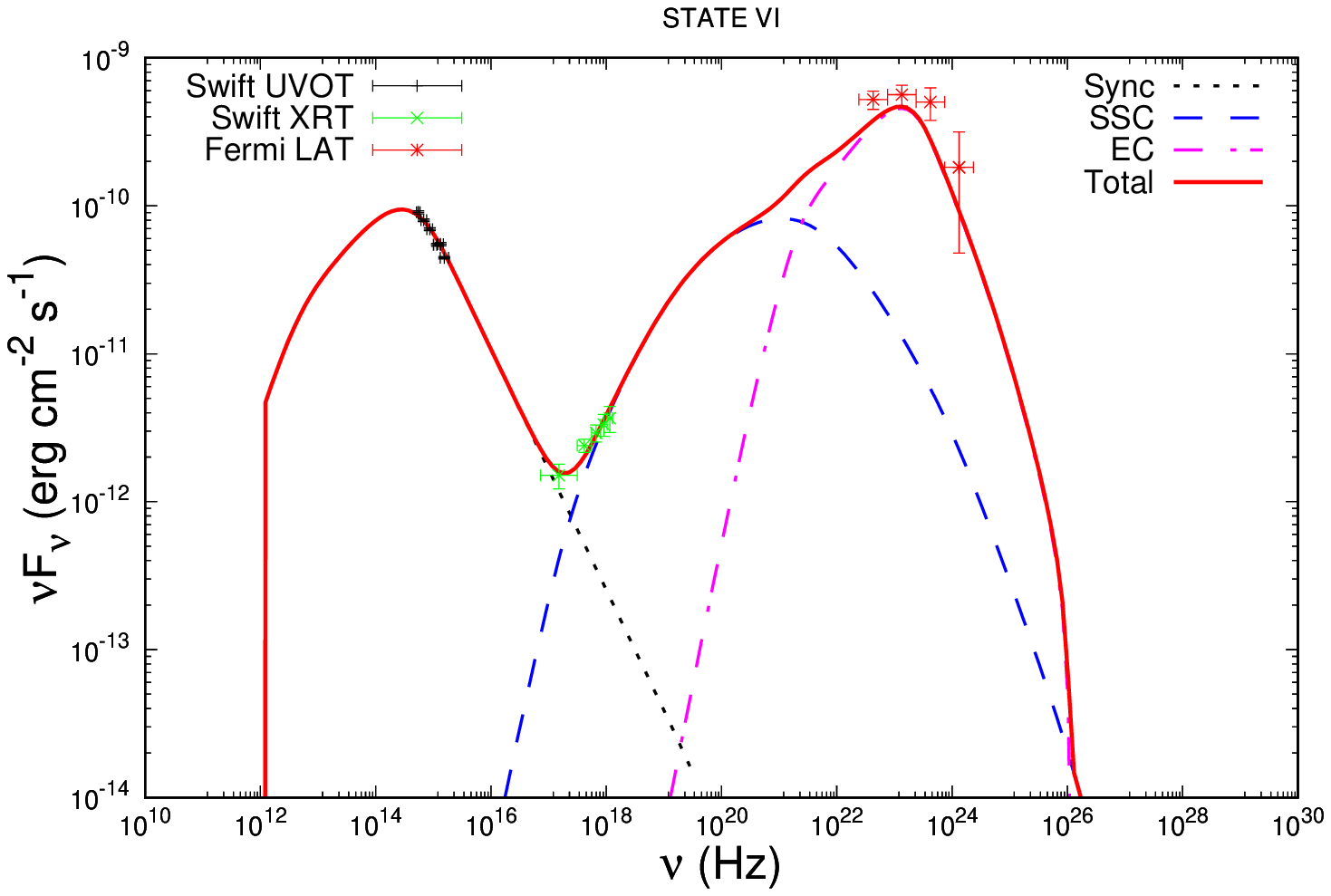}
    }   
\hbox{
\includegraphics[scale=0.24, angle=270]{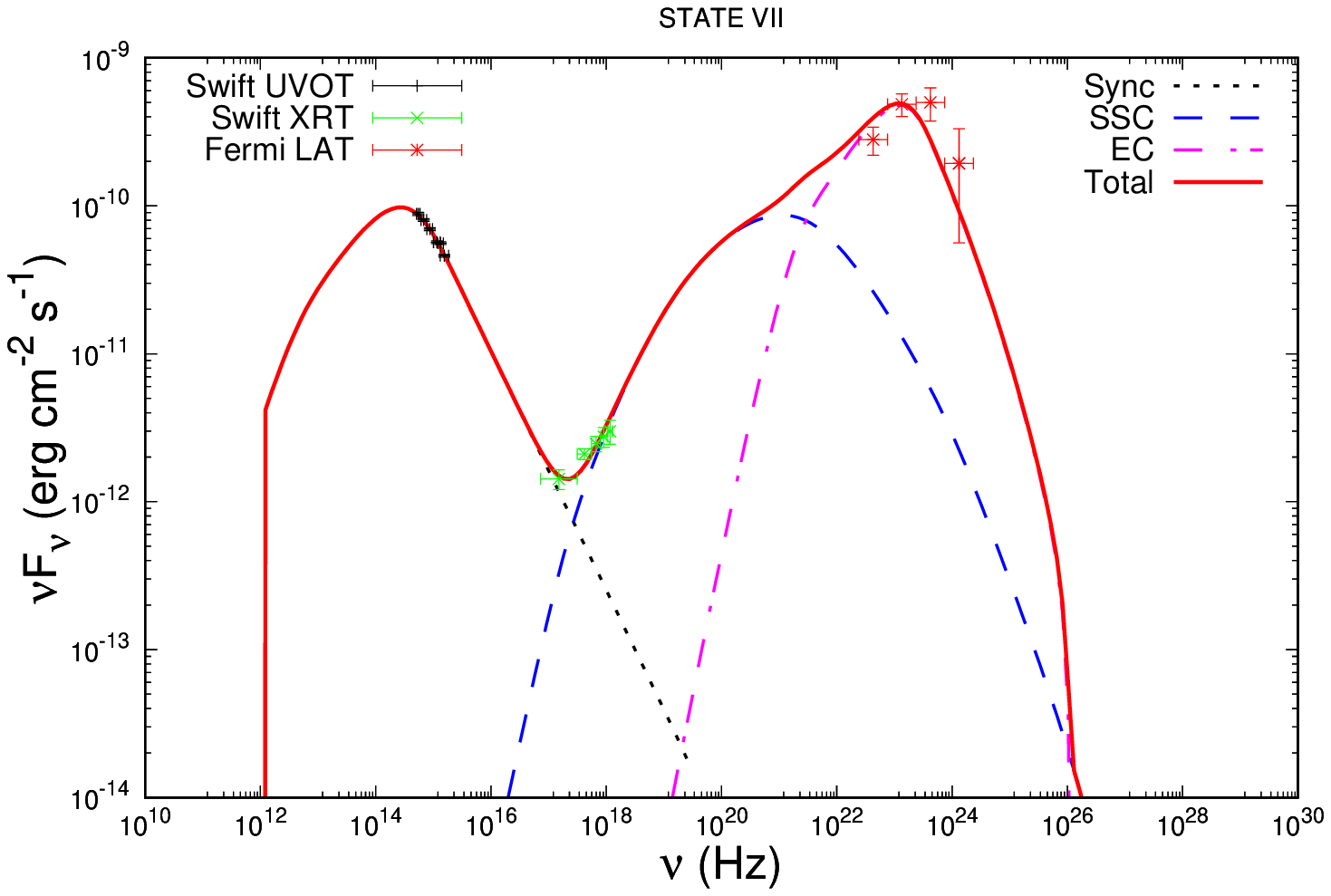}
\includegraphics[scale=0.24, angle=270]{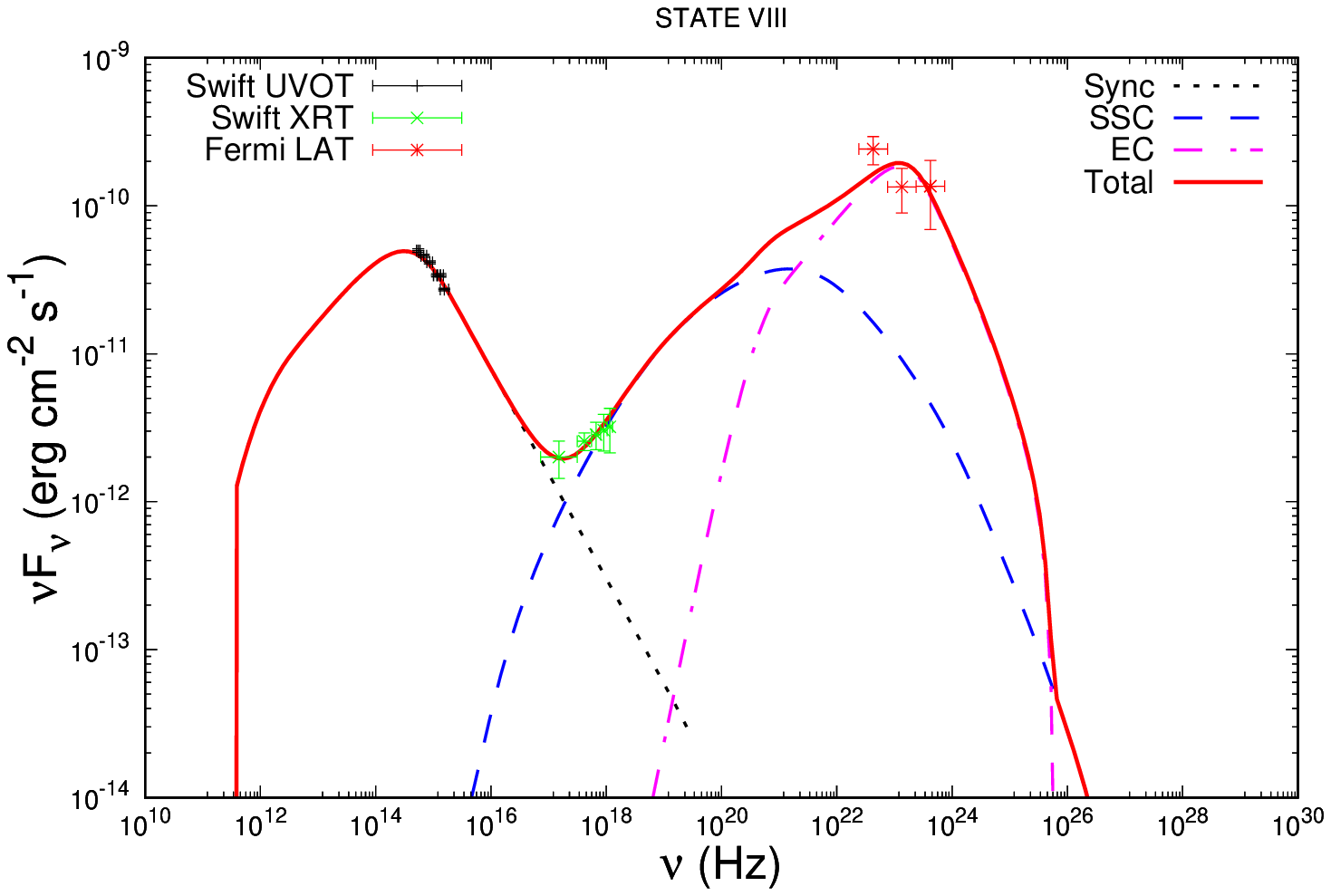}
\includegraphics[scale=0.24, angle=270]{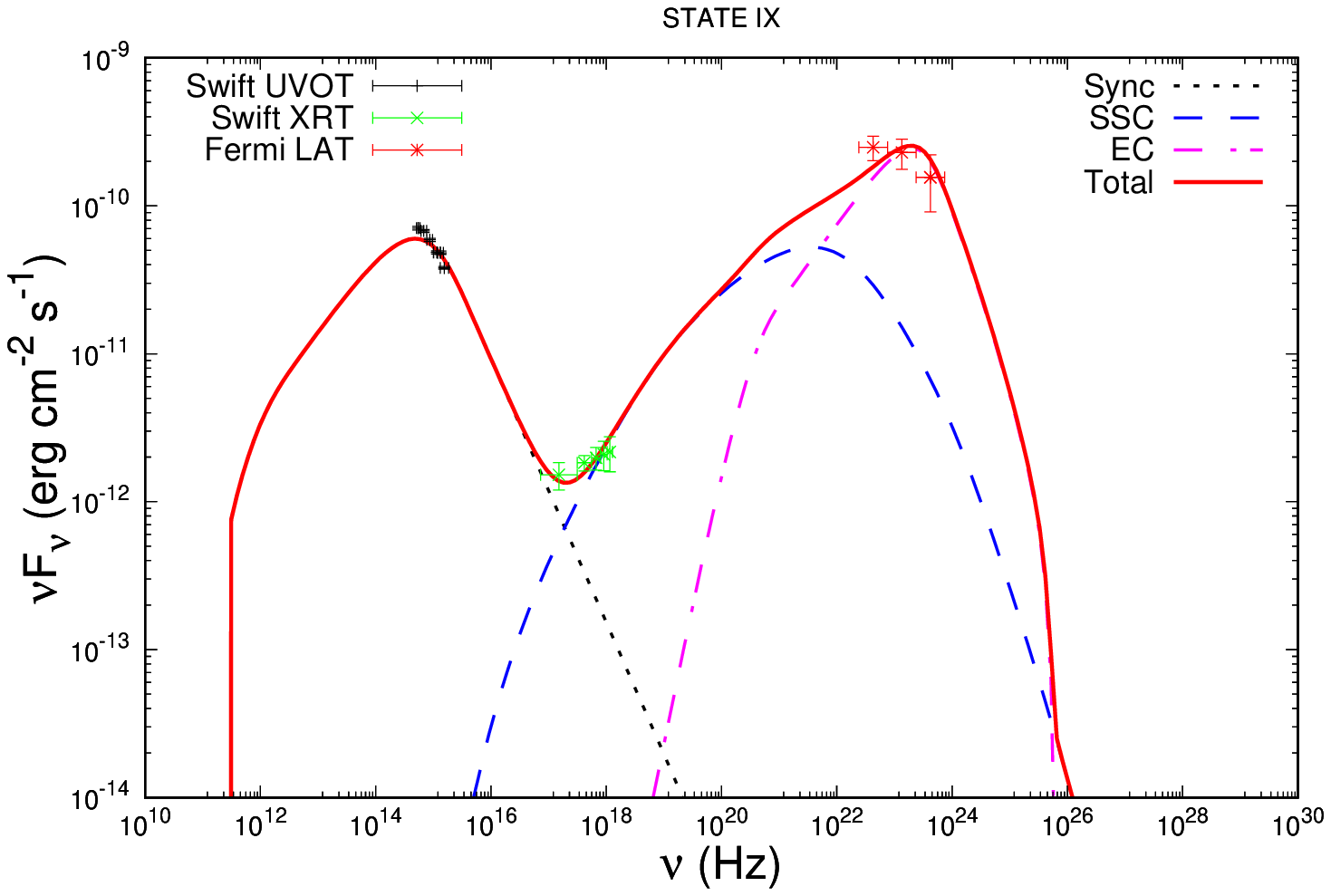}
    }
\hbox{
\includegraphics[scale=0.24, angle=270]{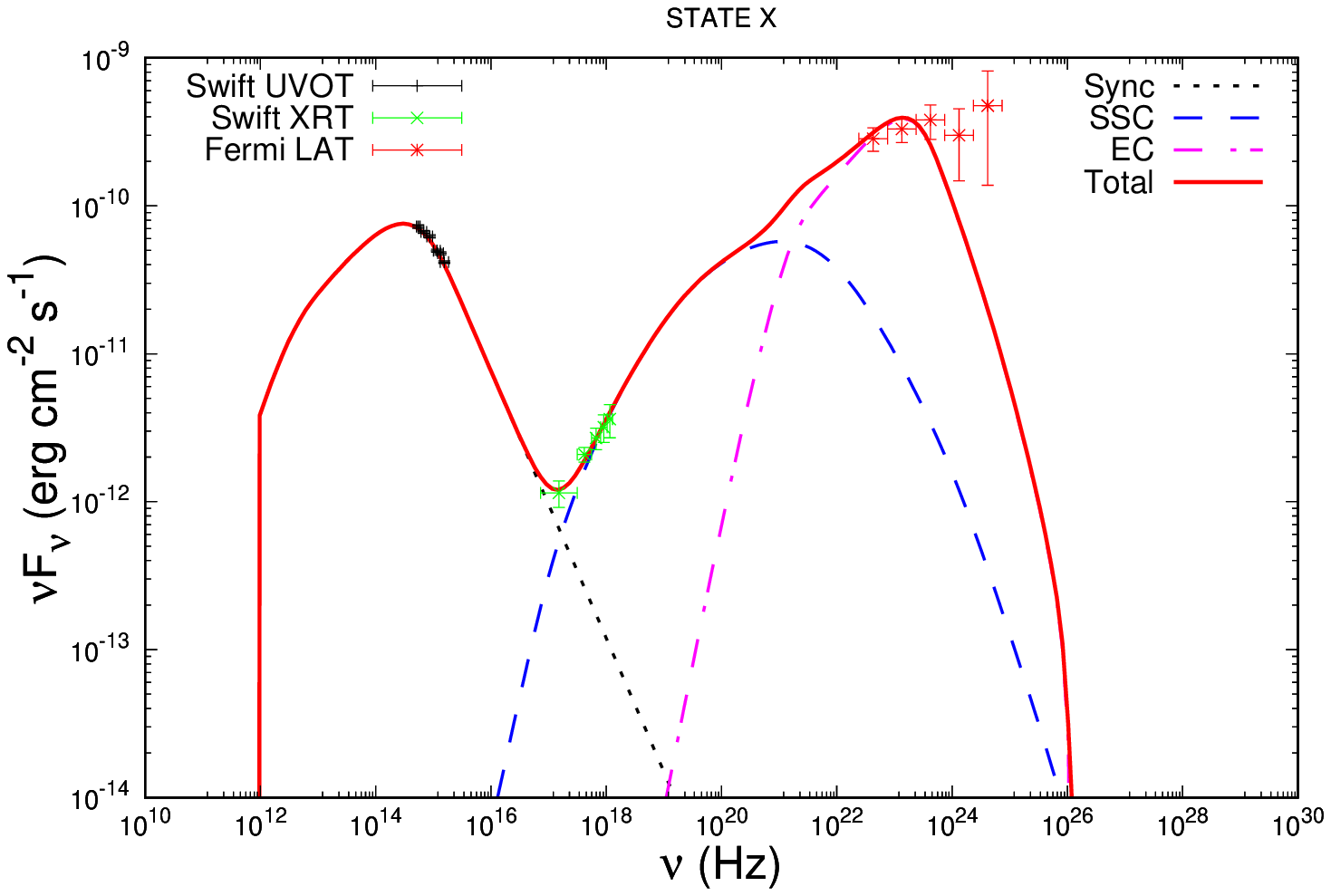}
}
}

\vspace{1cm}
\caption{Broadband SEDs of Ton\,599 for the ten selected flux intervals (see Table~\ref{tab:xrt_obs}). The solid red colour curves represent the total (Synchrotron, SSC, EC) SED fit.}
    \label{fig:sed}
\end{figure*}

\section{Summary and Discussion}
\label{diss}
The simultaneous multi-wavelength data at optical/UV, X-ray, and $\gamma$-ray energies enabled a thorough investigation into the temporal and spectral characteristics of Ton\,599 during the January 2023 flare.
The  $\gamma$-ray light curve exhibits a flaring activity during the period MJD 59943 -- 59974, we defined this period as an active state of the source. During the January 2023 period,  the daily averaged $\gamma$-ray light curve reveals the five flaring components. Figure~\ref{new_component} illustrates that the components C1, C2, and C3 are symmetric, while the component C4 is asymmetric. However, due to substantial error in the asymmetric parameter value of component C5, its profile is categorized as both symmetric and moderately asymmetric. The symmetric flare profile can be attributed to the time interval required for particles or radiation to traverse the emission region, which is determined by spatial and geometric scales \citep{2000ApJ...542L.105T, 2001ApJ...563..569T}. Moreover, the overlay of short-time events can give rise to a symmetric flare \citep{1999ApJS..120...95V}. The asymmetry in the flare profile can be attributed to rapid electron injection in an emission region followed by delayed escape or slow radiative cooling. Moreover, it can be also produced by injection of short-living energetic electrons and wider jet opening angles  \citep{2010ApJ...722..520A}. 

The temporal analysis of MWL LC of Ton\,599 shows simultaneous and correlated variability in all the energy bands, suggesting that the broadband emission is from the same emission region and electron distribution. The increased variability amplitude in $\gamma$-rays, as compared to X-rays and optical/UV, is in accordance with results from previous blazar investigations. \citep[e.g, ][]{2005ApJ...629..686Z, 2019MNRAS.484.3168S, 2022MNRAS.514.4259M}. Importantly, there is a similarity between the variability pattern observed in different energy bands and the shape of the SED. The broadband SED analysis reveals that the X-ray emission occurs before the break energy, indicating its origin from low-energy electrons. Contrarily, the $\gamma$-ray and Optical/UV emissions occur near or after the break energy, suggesting their association with high-energy electrons. As a result, we expect a greater degree of variability in the $\gamma$-ray energy compared to the X-ray energies. This is due to the faster cooling of high-energy electrons relative to their low-energy counterparts. The increase of flux variability with increasing energy reveals spectral variability in the source, as noted by \citet{2005ApJ...629..686Z}.

To acquire insights into the fluctuations of physical parameters responsible for the flux enhancement during the active state (MJD 59943-59974), we divided the MWL LC into ten flux intervals and subsequently, we carried out a detailed broadband spectral study of the source in these selected intervals. 
During the process of broadband SED fitting, we considered the condition of equipartition between the energy density of particles and the energy density of the magnetic field. We also assumed a steady state emission in the selected flux intervals.  In our modelling approach, we employed a BPL electron distribution, which has been widely used in previous studies on blazars \citep[e.g.][]{Sahayanathan_2018,2019MNRAS.484.3168S}. A BPL electron distribution, experiencing synchrotron, SSC, and EC losses, effectively replicates the broadband emission across all flux intervals. The synchrotron process is responsible for the low-energy emission in the optical/UV energy band, while the high-energy emissions in the X-ray and $\gamma$-ray band are associated with the SSC and EC processes. We found that in all of the flux intervals, the EC scattering of IR target photons provided a satisfactory fit to the data.
The resulting best-fit parameters \emph{viz}. \emph{p}, \emph{q}, \emph{B} and $\Gamma$ for different flux intervals are shown in Table~\ref{parameters}. The \emph{B} varies in the
the range of 1.52 -- 1.66,  which is larger than the average value of \emph{B} for FSRQs \citep{2015MNRAS.448.1060G} but falls within the physically acceptable range  \citep{2015MNRAS.454.1767L}. Furthermore, it increases with the increase in the flux (see Table~\ref{parameters}). The obtained spectral indices \emph{p} and \emph{q} vary between 1.80 -- 2.34 and 4.04 -- 4.92 respectively. The indices get harder as the source becomes brighter. These values of indices show that the steepness of the spectra after the break energy is greater than what would be expected from synchrotron cooling alone. This suggests that the break in the particle distribution is not solely due to cooling effects and indicates the need for an alternative explanation. One possibility is the presence of an energy-dependent diffusion coefficient. Previous studies, such as \citet{2007A&A...465..695Z}, have demonstrated that with an energy-dependent diffusion coefficient, the spectral cutoff exhibits a sub-exponential or steeper form at high energies.
 Moreover, in the diffuse shock particle acceleration process, the hardness or steepness of indices is determined by the shock speed, shock field obliquity, nature of particle scattering, and magnitude of turbulence \citep{2012ApJ...745...63S}.  The value of $\Gamma$ ranges from 28.25 to 37.15 in all the flux intervals, and there is no apparent trend observed in $\Gamma$ between the low and high flux emissions.  The source has also been studied earlier by various authors in different flux intervals. Notably, \citet{rajput2021gammaray} examined its 2021 flare, while \citet{Prince_2019} and \citet{2020MNRAS.492...72P} studied a high-activity state in 2017. Our best-fit parameters (\emph{B}, \emph{p} and \emph{q}) are in agreement with those of \citet{2020MNRAS.492...72P}, who however have fixed the value of $\Gamma$ at 15. The \citet{rajput2021gammaray} and \citet{Prince_2019} have focused mostly on the \emph{Fermi} spectrum of the source and haven't carried out the broadband spectral modelling of the source.
 Using the broadband observations \emph{viz}. optical/UV, X-ray and $\gamma$-ray of Ton\,599, we have investigated the changes in underlying physical parameters during a January 2023 active state of the source.  By incorporating additional data at the VHE  range and conducting the broadband analysis during various active states of the source,  a better constraint on the underlying physical parameters responsible for the flux variation in the source can be achieved in future research work.

\section{Acknowledgements}
AM, SS and NI are thankful to the Department of Atomic Energy (DAE), Board of Research in Nuclear Sciences (BRNS), Govt of India via Sanction Ref No.: 58/14/21/2019-BRNS for the financial support. ZS is supported by the Department of Science and Technology, Govt. of India, under the INSPIRE Faculty grant (DST/INSPIRE/04/2020/002319).
\section{Data Availability}
The multi-wavelength data used in this work is publicly available at the data browse section of the respective observatories. Codes used in this work will be shared on request to the corresponding authors (email: aqibmanzoor1111@gmail.com and shahzahir4@gmail.com)


\bibliographystyle{mnras}
\bibliography{refrence} 

\label{lastpage}
\end{document}